\documentclass[twocolumn,aps,prl,superscriptaddress]{revtex4}
\usepackage{graphicx}
\newcommand{\expval}[1]{\left\langle #1 \right\rangle}
\begin{document}
\title{Observation of Berry's Phase in a Solid State Qubit}
\author{P.~J.~Leek}
\affiliation{Department of Physics, ETH Z\"urich, CH-8093, Z\"urich,
Switzerland.}
\author{J.~M.~Fink}
\affiliation{Department of Physics, ETH Z\"urich, CH-8093, Z\"urich,
Switzerland.}
\author{A.~Blais}
\affiliation{D\'epartement de Physique, Universit\'e de
Sherbrooke, Sherbrooke, Qu\'ebec, J1K 2R1 Canada.}
\author{R.~Bianchetti}
\affiliation{Department of Physics, ETH Z\"urich, CH-8093, Z\"urich,
Switzerland.}
\author{M.~G\"oppl}
\affiliation{Department of Physics, ETH Z\"urich, CH-8093, Z\"urich,
Switzerland.}
\author{J.~M.~Gambetta}
\affiliation{Institute for Quantum Computing and Department of Physics and Astronomy,
University of Waterloo, Waterloo, Ontario, N2L 3G1 Canada.}
\affiliation{Departments of Applied Physics and Physics, Yale
University, New Haven, CT 06520, USA.}
\author{D.~I.~Schuster}
\affiliation{Departments of Applied Physics and Physics, Yale
University, New Haven, CT 06520, USA.}
\author{L.~Frunzio}
\affiliation{Departments of Applied Physics and Physics, Yale
University, New Haven, CT 06520, USA.}
\author{R.~J.~Schoelkopf}
\affiliation{Departments of Applied Physics and Physics, Yale
University, New Haven, CT 06520, USA.}
\author{A.~Wallraff}
\affiliation{Department of Physics, ETH Z\"urich, CH-8093, Z\"urich,
Switzerland.}
\date{\today}

\begin{abstract}
In quantum information science, the phase of a wavefunction plays an important role in encoding information.
While most experiments in this field rely on dynamic effects to manipulate this information,
an alternative approach is to use geometric phase, which has been argued to have potential fault tolerance.
We demonstrate the controlled accumulation of a geometric phase, Berry's phase, in a superconducting qubit,
manipulating the qubit geometrically using microwave radiation, and observing the accumulated phase in an interference experiment.
We find excellent agreement with Berry's predictions, and also observe a geometry dependent contribution to dephasing.
\end{abstract}

\maketitle

When a quantum mechanical system evolves cyclically in time such that it returns to its initial physical state, its wavefunction can acquire a geometric phase factor in addition to the familiar dynamic phase \cite{ShapereWilczek,Anandan1992}. If the cyclic change of the system is adiabatic, this additional factor is known as Berry's phase \cite{Berry1984}, and is, in contrast to dynamic phase, independent of energy and time.

In quantum information science \cite{NielsenChuang}, a prime goal is to utilize coherent control of quantum systems to process information, accessing a regime of computation unavailable in classical systems. Quantum logic gates based on geometric phases have been demonstrated in both nuclear magnetic resonance \cite{Jones2000} and ion trap based quantum information architectures \cite{Leibfried2003}. Superconducting circuits \cite{Wendin2006,Devoret2004} are a promising solid state platform for quantum information processing \cite{Yamamoto2003,Steffen2006a,Niskanen2007,Plantenberg2007,Majer2007,Sillanpaa2007}, in particular due to their potential scalability. Proposals for observation of geometric phase in superconducting circuits \cite{Falci2000,Wang2002,Blais2003,Peng2006,Mottonen2006} have existed since shortly after the first coherent quantum effects were demonstrated in these systems \cite{Nakamura1999}.

\begin{figure}[bp]
\includegraphics[width=0.9 \columnwidth]{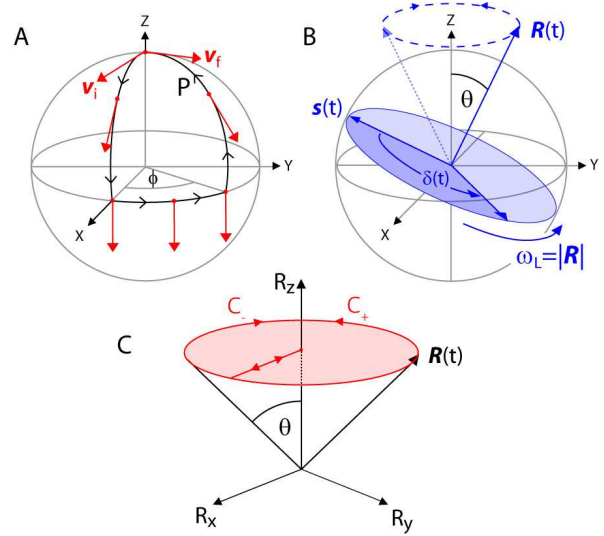}
\caption{({\bf A})~Parallel
transport of the vector $\textbf{\textit{v}}_{\rm{i}}$ on a spherical surface
around the closed path $P$ results in it rotating by a geometric
angle $\phi$ to $\textbf{\textit{v}}_{\rm{f}}$ when it returns to its initial
position. ({\bf B})~Dynamics of the Bloch
vector $\textbf{\textit{s}}$ of a qubit in the presence of a bias field
$\textbf{\textit{R}}$ tilted by an angle $\theta$ from the z-axis. ({\bf C})~Parameter
space of the Hamiltonian for the same case.}
\end{figure}

Geometric phases are closely linked to the classical concept of
parallel transport of a vector on a curved surface. Consider, for
example, a tangent vector $\textbf{\textit{v}}$ on the surface of a sphere being
transported from the North pole around the path $P$ shown in
Fig.~1A, with $\textbf{\textit{v}}$ pointing South at all
times. The final state of the vector $\textbf{\textit{v}}_{\rm{f}}$ is rotated
with respect to its initial state $\textbf{\textit{v}}_{\rm{i}}$ by an angle
$\phi$ equal to the solid angle subtended by the path $P$ at the origin.
Thus, this angle is dependent on the geometry of the path $P$,
and is independent of the rate at which it is traversed. As a result, departures from the original path that leave
the solid angle unchanged will not modify $\phi$. This robustness has been interpreted as a potential fault tolerance when applied to quantum information processing \cite{Jones2000}.

The analogy of the quantum geometric phase with the above classical picture is particularly clear
in the case of a two-level system (a qubit) in the presence of a
bias field that changes in time. A familiar example is a spin-1/2
particle in a changing magnetic field. The general Hamiltonian for
such a system is ${H}=\hbar\textbf{\textit{R}}\cdot\mbox{\boldmath$\sigma$}/2$,
where $\mbox{\boldmath$\sigma$}=( \sigma_{\rm{x}}, \sigma_{\rm{y}}, \sigma_{\rm{z}})$
are the Pauli operators, and $\textbf{\textit{R}}$ is the bias field vector,
expressed in units of angular frequency. The qubit dynamics is best
visualized in the Bloch sphere picture, in which the qubit state
$\textbf{s}$ continually precesses about the vector $\textbf{\textit{R}}$,
acquiring dynamic phase $\delta(t)$ at a rate $R=|\textbf{\textit{R}}|$ (see
Fig.~1B). When the direction of $\textbf{\textit{R}}$ is now changed adiabatically in
time (i.e. at a rate slower than $R$), the qubit additionally acquires Berry's phase, while remaining
in the same superposition of eigenstates
with respect to the quantization axis $\textbf{\textit{R}}$.
The path followed by $\textbf{\textit{R}}$ in the three dimensional
parameter space of the Hamiltonian (see Fig.~1C) is
the analogue of a path in real space in the classical case. When
$\textbf{\textit{R}}$ completes the closed circular path $C$, the geometric phase acquired by an eigenstate is
$\pm\Theta_C/2$ \cite{Berry1984}, where $\Theta_C$ is the solid
angle of the cone subtended by $C$ at the origin. The
$\pm$ sign refers to the opposite phases acquired by the ground or
excited state of the qubit, respectively. For the circular path
shown in Fig.~1C, the solid angle is given by
$\Theta_C=2\pi(1-\cos{\theta})$, depending only on the cone angle $\theta$.

We describe an experiment carried out on an individual
two-level system realized in a superconducting electronic circuit.
The qubit is a Cooper pair box \cite{Shnirman1997,Bouchiat1998} with an energy level
separation of $\hbar\omega_{\rm{a}}\approx h\times3.7~\rm{GHz}$ when
biased at charge degeneracy, where it is optimally protected from
charge noise \cite{Vion2002}. The qubit is embedded in a one-dimensional microwave
transmission line resonator with resonance frequency
$\omega_{\rm{r}}/2\pi \approx 5.4~\rm{GHz}$ (see
Fig.~2A). In this architecture, known as circuit
quantum electrodynamics (QED) \cite{Blais2004,Wallraff2004}, the qubit
is isolated effectively from its electromagnetic environment,
leading to a long energy relaxation time of $T_1 \approx
10~\mu\rm{s}$, and a spin-echo phase coherence time of
$T_2^{\rm{echo}} \approx 2~\mu\rm{s}$. In addition, the
architecture allows for a high visibility dispersive readout of the
qubit state \cite{Wallraff2005}.

Fast and accurate control of the bias field $\textbf{\textit{R}}$ for this
superconducting qubit is achieved through phase and amplitude
modulation of microwave radiation coupled to the qubit through the
input port of the resonator (see Fig.~2A). The qubit
Hamiltonian in the presence of such radiation is
\[ {H}=\frac{\hbar}{2}\omega_{\rm{a}} {\sigma}_{\rm{z}}+\hbar\Omega_{\rm{R}}\cos{(\omega_{\rm{b}}t+\varphi_R)} {\sigma}_{\rm{x}},\]
where $\hbar\Omega_{\rm{R}}$ is the dipole interaction strength between the qubit and a microwave field of frequency $\omega_{\rm{b}}$ and phase $\varphi_R$. Thus $\Omega_{\rm{R}}/2\pi$ is the Rabi frequency that results from resonant driving.
The above Hamiltonian may be transformed to a frame rotating at the frequency $\omega_{\rm{b}}$ using the unitary transformation ${H}'=UHU^{-1}-i\hbar U{\dot{U}}^{-1}$, where $U=e^{i\omega_b t\sigma_z/2}$. Ignoring terms oscillating at $2\omega_b$ (the rotating wave approximation), the transformed Hamiltonian takes the form \[ H'\approx\frac{\hbar}{2}(\Delta {\sigma}_{\rm{z}}+\Omega_{\rm{x}} {\sigma}_{\rm{x}}+\Omega_{\rm{y}} {\sigma}_{\rm{y}}),\]
where $\Omega_{\rm{x}}=\Omega_{\rm{R}}\cos\varphi_R$ and $\Omega_{\rm{y}}=\Omega_{\rm{R}}\sin\varphi_R$. This is equivalent to the generic situation depicted in
Fig.~1B and C, with $\textbf{\textit{R}}=(\Omega_{\rm{x}},\Omega_{\rm{y}},\Delta)$ and
$\Delta=\omega_{\rm{a}}-\omega_{\rm{b}}$ being the detuning between
the qubit transition frequency and the applied microwave frequency.
In our experiment we keep $\Delta$ fixed, and control the bias field to trace
circular paths of different radius $\Omega_{\rm{R}}$.

We measure Berry's phase in a Ramsey fringe interference experiment
by initially preparing an equal superposition of the qubit ground
and excited states, which acquires a relative geometric phase $\gamma_C=
2\pi(1-\cos{\theta})$, equal to the total solid angle
enclosed by the cone depicted in Fig.~1C, with
$\cos\theta=\Delta/(\Omega_{\rm{R}}^2+\Delta^2)^{1/2}$. As the bias field adiabatically follows
the closed path $C_\pm$, the qubit state acquires both a dynamic
phase $\delta(t)$ and a geometric phase $\gamma_C$, corresponding to
a total accumulated phase $\phi=\delta(t)\pm\gamma_C$ (the $\pm$
sign denoting path direction) which we extract by performing full
qubit state tomography \cite{NielsenChuang}. To directly observe only the
geometric contribution, we use a spin echo \cite{Abragam}
pulse sequence that cancels the dynamic phase as explained below.

\begin{figure}[b]
\includegraphics[width=1. \columnwidth]{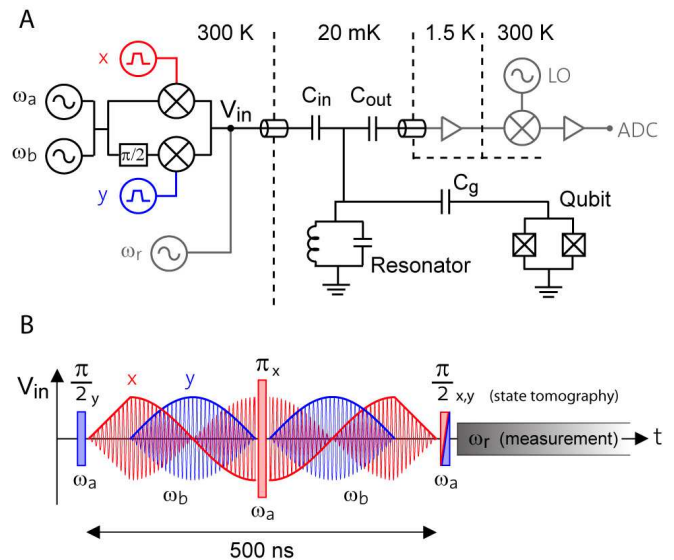}
\caption{({\bf A})~Simplified
circuit diagram of the experimental setup. In the center at
$20~\rm{mK}$ is the resonator/qubit system, with the resonator
represented by a parallel LC circuit, and the qubit, a split Cooper
pair box, capacitively coupled to the resonator
through $C_{\rm{g}}$. The resonator is coupled to input and output
transmission lines via the capacitors $C_{\rm{in}}$ and
$C_{\rm{out}}$. Three different pulse-modulated microwave frequency
signals are applied to the resonator input. The two signals required
for qubit manipulation, one at the qubit transition frequency
$\omega_{\rm{a}}/2\pi$, and a detuned signal $\omega_{\rm{b}}/2\pi$
are modulated using mixers to the pattern shown in (B). The
signal at the resonator frequency $\omega_{\rm{r}}/2\pi$, used to
measure the qubit state, is turned on after the
pulse sequence is applied. ({\bf B})~Schematic pulse sequence for the case $n=0.5$.
Resonant pulses, shown as shaded rectangles, are $12~\rm{ns}$ in
length. The two quadrature bias microwave fields (x: red, y: blue)
are represented as sinusoids with modulation amplitude shown by
solid lines. The linear ramps at the start and end of these sections
correspond to moving adiabatically between the $\Omega_{\rm{R}}=0$
axis and the $\Omega_{\rm{R}}=\rm{const.}$ circle (Fig.~1C).}
\end{figure}

The complete sequence (see Fig.~2B) starts by
preparing the initial $\sigma_z$ superposition state with a resonant
$\pi$/2 pulse. Then the path $C_-$ is traversed, causing the qubit
to acquire a phase $\phi_{\rm{-}}=\delta(t)-\gamma_C$. Applying a
resonant spin echo $\pi$ pulse to the qubit about an orthogonal axis
now inverts the qubit state, effectively inverting the phase
$\phi_{\rm{-}}$. Traversing the control field path again, but in the
opposite direction $C_+$, adds an additional phase
$\phi_{\rm{+}}=\delta(t)+\gamma_C$. This results in total in a
purely geometric phase $\phi=\phi_{\rm{+}}-\phi_{\rm{-}}=2\gamma_C$
being acquired during the complete sequence which we denote as
$C_{-+}$. Note that unlike the geometric phase, the dynamic phase is
insensitive to the path direction, and is hence completely canceled. At the end of the
sequence, we extract the phase of the qubit state using quantum
state tomography. In our measurement technique \cite{Wallraff2005} the
$z$ component of the qubit Bloch vector
$\expval{ \sigma_{\rm{z}}}$ is determined by measuring the
excited state population
$p_{\rm{e}}=(\expval{ \sigma_{\rm{z}}}+1)/2$. To extract the x
and y components, a resonant $\pi/2$ pulse rotating the qubit about
either the $x$ or $y$ axis is applied, and then the measurement is
performed, revealing $\expval{ \sigma_{\rm{y}}}$ and
$\expval{ \sigma_{\rm{x}}}$ respectively. The phase of the
quantum state after application of the control sequence is then
extracted as
$\phi=\tan^{-1}{(\expval{ {\sigma}_{\rm{y}}}/\expval{ {\sigma}_{\rm{x}}})}$.

In Fig.~3A we show the measured phase $\phi$ and its
dependence on the solid angle of the path, for a number of
different experiments, all carried out at
$\Delta/2\pi\approx50~\rm{MHz}$, and total pulse sequence time
$T=500~\rm{ns}$. Three parameters are varied; the path radius
$\Omega_{\rm{R}}$ (upper x-axis), the number of circular loops
traversed in each half of the spin echo sequence, $n$, and the direction of
traversal of the paths ($C_{-+}$ and $C_{+-}$). The measured phase
is in all cases seen to be linear in solid angle as
$\Omega_{\rm{R}}$ is swept, with a root-mean-squared deviation across all data sets of $0.14~\rm{rad}$ from the expected lines of
slope $2n$. Thus, all results are in close agreement with the
predicted Berry's phase and it is clearly demonstrated that we are
able to accurately control the amount of phase accumulated
geometrically. Also we note that the dynamic phase is indeed
effectively eliminated by the spin echo. Reversing the overall
direction of the paths is observed to invert the sign of the phase (Fig.~3A).
Traversing the circular paths on either side of the spin echo pulse
in the same direction ($C_{++}$) as a control experiment is shown to
result in zero measured phase (Fig.~3A).

\begin{figure}[tp!]
\includegraphics[width=0.85 \columnwidth]{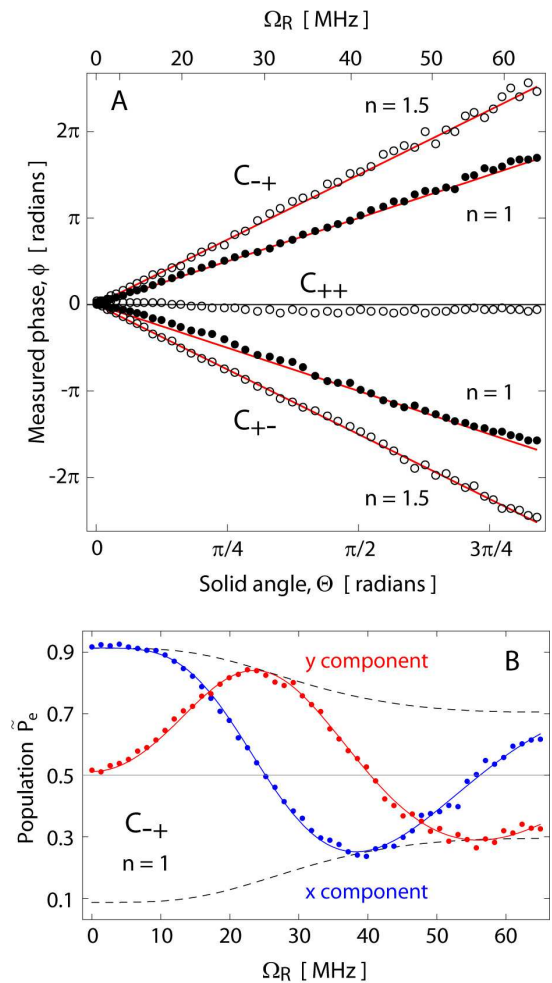}
\caption{({\bf A})~Measured phase $\phi$ versus solid angle $\Theta$ of a single conical path (lower axis).
The applied microwave field amplitude is indicated on the upper axis (in units of the induced
Rabi frequency $\Omega_{R}$ for resonant driving).
Closed circles correspond to experiments in which $n=1$ circular paths are traversed during each half of the spin echo sequence,
and filled circles to the case $n=1.5$. Subscripts $\pm$ of labels $C_{\pm\pm}$ correspond to
the path direction before and after the spin echo $\pi$ pulse. Red
solid lines are of slope $n=\pm 1$, $\pm 1.5$. The $C_{++}$ experiment was carried out with $n=1.5$.
({\bf B})~State tomography data for the $C_{-+}$ experiment with $n=1$. Plotted
is the qubit excited state population after tomography pulses to
extract $\expval{ \sigma_{\rm{x}}}$ (blue, $p_{\rm{e}}=(\expval{ \sigma_{\rm{x}}}+1)/2$) and
$\expval{ \sigma_{\rm{y}}}$ (red, $p_{\rm{e}}=(\expval{ \sigma_{\rm{y}}}+1)/2$). Lines are fits to
Berry's phase, with a geometric dephasing envelope function (dashed
lines, described in the text and Fig.~4). In all cases, the total
pulse sequence time is $T=500~\rm{ns}$, and the detuning is
$\Delta/2\pi\approx50~\rm{MHz}$. To accumulate measurement
statistics sequences are repeated $2\times10^5$ times.}
\end{figure}

To observe a pure Berry's phase, the rate of rotation of the bias
field direction must be much less than the Larmor rate $R$ of the
qubit in the rotating frame, to ensure adiabatic qubit dynamics.
For the case of constant cone angle $\theta$, this translates to
the requirement $A=\dot\varphi_R\sin\theta/2R\ll1$. If the
Hamiltonian is changed non-adiabatically, the qubit state can no longer exactly follow the
effective field $\textbf{\textit{R}}$, and the geometric phase acquired deviates from Berry's phase \cite{Aharonov1987}.
For the experiments here, $A\le0.04$, and deviation of the measured phase
from Berry's phase is not discernable. We have also verified experimentally that in this adiabatic limit
the observed phase is independent of the total sequence time $T$.

In Fig.~3B, a measurement of the x and y components of
the qubit state from which the Berry's phase is extracted is shown.
Interestingly, the visibility of the observed interference pattern is seen
to have a dependence on $\Omega_{\rm{R}}$. Since the data is
taken at fixed total sequence time, this is not due to conventional $T_2$ dephasing,
which is also independently observable as a function of time, but
can be explained as due to geometric dephasing, an effect
dependent on the geometry of the path \cite{De2003}.

\begin{figure}[tp!]
\includegraphics[width=0.85 \columnwidth]{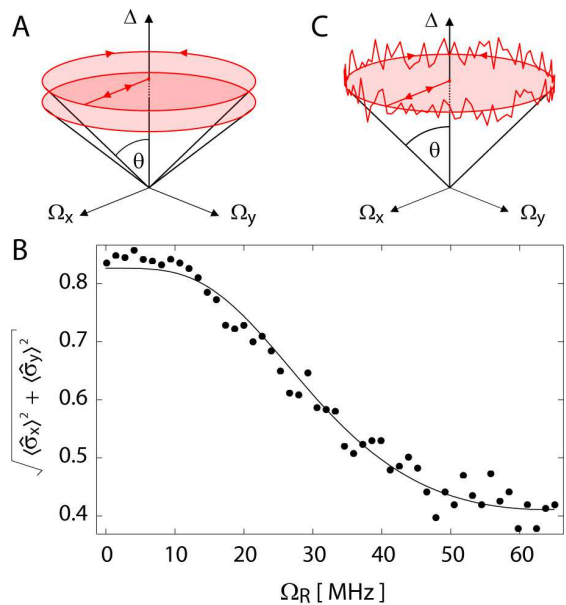}
\caption{({\bf A})~Low frequency fluctuations in
$\Delta$ change the solid angle enclosed by the path from one
measurement to the next, and cause geometric dephasing with a
characteristic dependence on the cone angle and bias field
amplitude. ({\bf B})~Magnitude of the equatorial component of the
Bloch vector $(\expval{ {\sigma}_{\rm{x}}}^2+\expval{ {\sigma}_{\rm{y}}}^2)^{1/2}$
for the data shown in Fig.~3B, plotted as a function
of drive amplitude $\Omega_{\rm{R}}$. The fit is to a geometric
dephasing factor $e^{-\sigma_\gamma^2/2}$ where $\sigma_\gamma^2$ is the variance of the geometric phase.
({\bf C})~The conical parameter space path in the presence of high frequency
($f\gg T^{-1}$) noise in $\Delta$, having no effect on the total solid angle.}
\end{figure}

In our experiment, dephasing is dominated by low frequency
fluctuations in the qubit transition frequency $\omega_{\rm{a}}$
(and thus $\Delta$) induced by charge noise coupling to the qubit
\cite{Ithier2005}. The spin echo pulse sequence effectively cancels the dynamic
dephasing due to the low frequency noise. However, the geometric
phase is sensitive to slow fluctuations, since they cause the solid
angle subtended by the path at the origin to change from one
measurement to the next (see Fig.~4A). The effect on the
geometric phase of such fluctuations in the classical control
parameters of the system has been studied theoretically \cite{De2003}.
In the limit of the fluctuations being slower than the time scale of the spin
echo sequence, the variance of the geometric phase $\sigma_\gamma^2$ has itself a
purely geometric dependence, $\sigma_\gamma^2=\sigma_\omega^2(2n\pi\sin^2\theta/R)^2$,
where $\sigma_\omega^2$ is the variance of the fluctuations in
$\omega_{\rm{a}}$ \cite{De2003}. In Fig.~4B, we show the
observed dependence of the coherence on geometry explicitly by plotting
$(\expval{{\sigma}_{\rm{x}}}^2+\expval{ {\sigma}_{\rm{y}}}^2)^{1/2}$
versus $\Omega_{\rm{R}}$, which fits well to the expected dependence
$e^{-\sigma_\gamma^2/2}$. This is also in agreement with the raw data in Fig.~3B.

Hence we have observed an important geometric contribution to
dephasing that occurs when geometric operations are carried out in
the presence of low frequency fluctuations. In contrast, higher
frequency noise in $\omega_{\rm{a}}$ is expected to have little
influence on Berry's phase (provided adiabaticity is maintained), since its effect on the solid
angle is averaged out (see Fig.~4C). This
characteristic robustness of geometric phases to high frequency
noise may be exploitable in the realization of logic gates for
quantum computation, although the effect of geometric
dephasing due to low frequency noise must be taken into account.

\begin{acknowledgments}
We thank P. Maurer and L. Steffen for their contributions to the project,
and A. Shnirman, J. Blatter, G. De Chiara and G. M. Palma for valuable discussions.
This work was supported by the Swiss National Science Foundation and by ETH Z\"urich.
P.J.L. acknowledges support from the EC via an Intra-European Marie-Curie Fellowship.
A.B. was supported by NSERC, CIFAR and FQRNT.
D.I.S., L.F. and R.J.S. acknowledge support from the National Security Agency under the Army Research Office, the NSF and Yale University. L.F. acknowledges partial support from the CNR-Istituto di Cibernetica, Pozzuoli, Italy.
J.M.G. was partially supported by ORDCF and MITACS.
\end{acknowledgments}
\bibliographystyle{apsrev}

\end{document}